\documentclass[runningheads,a4paper]{llncs}
\pdfoutput=1
\usepackage{amssymb}
\usepackage{graphicx}
\usepackage{url}

\usepackage{listings}
\usepackage{color}
\usepackage{courier}
\usepackage{xspace}
\usepackage{caption}

\newcommand{\keywords}[1]{\par\addvspace\baselineskip
\noindent\keywordname\enspace\ignorespaces#1}

\newcommand\polymake{\texttt{poly\-make}\xspace}
\newcommand\RNG{\texttt{RNG}\xspace}
\newcommand{\R}{\mathbb{R}}
\newcommand{\Q}{\mathbb{Q}}
\newcommand\conv{{\rm conv}}

\begin{document}

\mainmatter
%================================================

%==== FILL IN ====================================
\title{The \polymake XML file format}  % Full title
\titlerunning{The \polymake XML file format} % Shor ttitle
\author{Ewgenij Gawrilow\inst{1} \and Simon Hampe\inst{2} \and Michael Joswig\inst{2}}
\authorrunning{Gawrilow, Hampe \& Joswig}
\institute{
TomTom International BV\\
\email{egawrilow@gmail.com}
%\url{Eugens.URL???}
\and
TU Berlin, Germany\\
\email{\{hampe,joswig\}@math.tu-berlin.de}}
%\url{http://page.math.tu-berlin.de/~{hampe,joswig}

\maketitle

\definecolor{mainkeycolor}{rgb}{0,0,1}
\definecolor{attrcolor}{RGB}{140,75,0}
\definecolor{stringcolor}{RGB}{103,5,172}
\definecolor{nodenamecolor}{RGB}{0,100,14}

\lstdefinelanguage{pmschema}
{
  basicstyle=\footnotesize\ttfamily,
  keywords=[1]{attribute,default,datatypes,div,element,empty,external,grammar,include,inherit,list,mixed,namespace,notAllowed,parent,start,text,token},
  morekeywords=[2]{xsd, string, nonNegativeInteger, hexBinary},
  morekeywords=[3]{TopObject, SubObject, SimpleName, PropertyName, TopAttribs, ObjectContent, Property, LooseData, PropertyData, Attachment, AttachmentData, Text, Complex, VectorContents, ElementIndex, IdReference, Vector, MatrixContents, Matrix, TupleContents, Tuple},
  stringstyle=\color{stringcolor},
  keywordstyle=[1]\color{mainkeycolor},
  keywordstyle=[2]\color{attrcolor},
  keywordstyle=[3]\color{nodenamecolor},
  numberstyle=\tiny\ttfamily\color{black},
  stepnumber=1,
  numbers=left,
  captionpos=b,
  showspaces=false,
  showstringspaces=false,
  morestring=[b]",
}

\lstdefinelanguage{pmxml}
{
  basicstyle=\footnotesize\ttfamily,
  keywords=[1]{xml,pm, property, object, description,m,v,e,t,attachment},
  morekeywords=[2]{name, type, value, chk, version, encoding,i,cols,xmlns,CDATA},
  stringstyle=\color{stringcolor},
  keywordstyle=[1]\color{mainkeycolor},
  keywordstyle=[2]\color{attrcolor},
  numberstyle=\tiny\ttfamily\color{black},
  stepnumber=1,
  numbers=left,
  captionpos=b,
  escapechar=|,
  showspaces=false,
  showstringspaces=false,
  morestring=[b]",
}

\captionsetup[table]{skip=5pt}

\newcommand\keyone[1]{{\color{mainkeycolor}\texttt{#1}}}
\newcommand\keytwo[1]{{\color{attrcolor}\texttt{#1}}}
\newcommand\keythree[1]{{\color{nodenamecolor}\texttt{#1}}}

\begin{abstract}
We describe an XML file format for storing data from computations in algebra and geometry.
We also present a formal specification based on a RELAX-NG schema.
\keywords{XML; RELAX-NG; polymake}
\end{abstract}

\lstset{
  frame=single,
}

%------------------------------------------------------------
\section{Introduction}

\polymake is an open source software sytem for computing with a wide range of objects from polyhedral geometry and related areas \cite{DMV:polymake}.
This includes convex polytopes and polyhedral fans as well as matroids, finite permutation groups and ideals in polynomial rings.
As a key feature \polymake is designed as an extensible system, where each new version comes with new objects and new data types.
It is crucial to be able to store these objects in files since they themselves or part of the information on them result from costly computations.
The purpose of this note is to explain the general concept for \polymake's file format which is powerful enough to be able to grow with extensions to the software.

It is safe to say that the Extensible Markup Language (XML) is the de facto standard for exchanging data across platform and implementation boundaries.
%At first sight it may not be obvious what the true strength of this concept actually is.
XML imposes a tree structure on any kind of text, and it comes with a large array of tools which allow to process an XML file independent from the software which generated that file.
The tree structure makes it easy to ignore part of the data on input without losing consistence by pruning of subtrees.
Part of the realm of XML tools are transformation style sheets (XSLT) which, e.g., allow for simplified versioning or even translating into non-XML documents.
This makes XML especially useful for the long-term storage of data; see also \cite[\S1.1]{DGD-Gallery}.

XML file formats for storing mathematical content are ubiquitous.
The most widely used is MathML \cite{www:MathML} whose initial purpose was the presentation of mathematics in web pages.
However, by now there is also Content MathML and OpenMath \cite{www:OpenMath} which focus on the semantics.
Our goal here is to describe a simple XML format which is useful for the serialization of data which occur in computations in algebraic and polyhedral geometry.

%------------------------------------------------------------
\section{The file format by example}

We start out by looking at one short \polymake example XML file which stores a square and some of its properties, including a triangulation.
The formal description in terms of RELAX-NG \cite{spec:RELAXNG} is deferred until Section~\ref{sec:relax-ng} below.
\begin{lstlisting}[language=pmxml,caption={A \polymake XML file, encoding a square.},label=lstexample]
<?xml version="1.0" encoding="utf-8"?>
<?pm chk="56e977e8"?>
<object name="square" type="polytope::Polytope&lt;Rational&gt;"
    version="3.0" 
    xmlns="http://www.math.tu-berlin.de/polymake/#3">
  <description><![CDATA[cube of dimension 2]]></description>
  <property name="VERTICES">
    <m>
      <v>1 0 0</v>
      <v>1 1/3 0</v>
      <v>1 0 1/3</v>
      <v>1 1/3 1/3</v>
    </m>
  </property>
  <property name="FACETS" 
      type="SparseMatrix&lt;Rational,NonSymmetric&gt;">
    <m cols="3">
      <v> <e i="1">1</e> </v>
      <v> <e i="0">1/3</e> <e i="1">-1</e> </v>
      <v> <e i="2">1</e> </v>
      <v> <e i="0">1/3</e> <e i="2">-1</e> </v>
    </m>
  </property>
  <property name="LINEALITY_SPACE"><m /></property>
  <property name="BOUNDED" value="true" />
  <property name="N_FACETS" value="4" />
  <property name="N_VERTICES" value="4" />
  <property name="VOLUME" value="1/9" />
  <property name="TRIANGULATION">
    <object name="unnamed#0">
      <property name="FACETS">
        <m>
          <v>0 1 2</v>
          <v>1 2 3</v>
        </m>
      </property>
      <property name="F_VECTOR">
        <v>4 5 2</v>
      </property>
    </object>
  </property>
</object>
\end{lstlisting}

\paragraph{Mathematical background.}

A \emph{(convex) polytope} is the convex hull of finitely many points in a Euclidean space or, equivalently, the bounded intersection of finitely many affine halfspaces.
In \polymake points are encoded in terms of \emph{homogeneous coordinates} to allow for a consistent treatment of both polytopes and polyhedral cones.
Therefore, the polytope $\conv(S)$ for $S\subset \R^n$ finite is encoded as the cone spanned by $\{1\} \times S \subseteq \R \times \R^n$.
That is, in practical terms, the coordinates of points are always prepended with a $1$.
In our example, we are considering the unit square scaled by $1/3$.
Its \emph{vertices}, which form the unique generating set which is minimal with respect to inclusion, are written as $(1,0,0),(1,1/3,0),(1,0,1/3),(1,1/3,1/3)$.
Linear inequalities are encoded in a similar fashion.
The vector $(a_0,a_1,\dots,a_n)$ ought to be read as $a_0 + a_1 x_1 + \dots a_n x_n \geq 0$.
In this way, a point $p$ given in homogeneous coordinates fulfills an inequality given by a vector $a$, if and only if the scalar product $p \cdot a$ is nonnegative.
See \cite{Polyhedral+and+Algebraic+Methods} for an introduction to polytope theory from an algorithmic point of view.

\bigskip

Now we will walk the user through the Listing \ref{lstexample} line by line.

\paragraph{The parent object (Lines 3--6).}

% \begin{lstlisting}[language=pmxml,firstnumber=3]
% <object name="square" type="polytope::Polytope&lt;Rational&gt;" 
%     version="3.0" 
%     xmlns="http://www.math.tu-berlin.de/polymake/#3">
%   <description><![CDATA[cube of dimension 2]]></description>
% \end{lstlisting}

The mathematical entities relevant to \polymake occur as \emph{objects} each of which has a \emph{type}.
It will tell the parser what properties to expect and how to interpret them.
In this case the type describes a convex polytope with rational coordinates.
The \emph{version number} refers to a specific \polymake version.
Via XSLT this allows for automatic updates from one object or file format version to the next.
Optional names and descriptions provide additional human-readable information for quick identification.

\paragraph{Properties and matrices (Lines 7--24).}
% \begin{lstlisting}[language=pmxml,firstnumber=7]
%   <property name="VERTICES">
%     <m>
%       <v>1 0 0</v>
%       <v>1 1/3 0</v>
%       <v>1 0 1/3</v>
%       <v>1 1/3 1/3</v>
%     </m>
%   </property>
%   <property name="FACETS" 
%       type="SparseMatrix&lt;Rational,NonSymmetric&gt;">
%     <m cols="3">
%       <v> <e i="1">1</e> </v>
%       <v> <e i="0">1/3</e> <e i="1">-1</e> </v>
%       <v> <e i="2">1</e> </v>
%       <v> <e i="0">1/3</e> <e i="2">-1</e> </v>
%     </m>
%   </property>
%   <property name="LINEALITY_SPACE"><m /></property>
% \end{lstlisting}
Every object is made up of various \emph{properties}, which are identified by their names; their types are implicit.
\polymake keeps track of the type of each property.
The combination of the property's name with the version number (see above) uniquely determines the type. 

However, properties may also be encoded in a more involved way.
In our example the property named \texttt{FACETS} comes with the type \texttt{SparseMatrix} explicitly given.
This can be useful for saving space.
In general, it is legal to specify types which can be converted into the defined type of a property.
For sparse data types that conversion is only implicit, i.e., the matrix is never expanded into a dense matrix.
Most of the time the user will not notice the difference.

Any \polymake matrix is stored as a sequence of row vectors.
If it is sparse only the nonzero entries are written down.
The column of each entry is encoded in the \texttt{i} attribute, and the \texttt{cols} attribute of the matrix indicates the total number of columns of the matrix.
In this case property \texttt{FACETS} encodes the matrix with the row vectors
$(0,1,0)$, $(1/3, -1,0)$, $(0,0,1)$, $(1/3,0,-1)$,
and this yields the non-redundant exterior description
\[ x \geq 0, \quad x \leq 1/3, \quad y \geq 0, \quad y \leq 1/3 \enspace .\]
If \polymake encounters a property with an unknown name, that property is discarded --- but a backup file is created.
% See also the Section \ref{attachments} on attachments for flexible storage of additional arbitrary data.

\paragraph{Primitive properties (Lines 25--28).}
% \begin{lstlisting}[language=pmxml,firstnumber=25]
%   <property name="BOUNDED" value="true" />
%   <property name="N_FACETS" value="4" />
%   <property name="N_VERTICES" value="4" />
%   <property name="VOLUME" value="1/9" />
% \end{lstlisting}
Simple properties containing, e.g., numbers (integer, rational or float) or boolean values are stored in an XML attribute named \texttt{value}.

\paragraph{Subobjects (Lines 29--41).}
% \begin{lstlisting}[language=pmxml,firstnumber=29]
%   <property name="TRIANGULATION">
%     <object name="unnamed#0">
%       <property name="FACETS">
%         <m>
%           <v>0 1 2</v>
%           <v>1 2 3</v>
%         </m>
%       </property>
%       <property name="F_VECTOR">
%         <v>4 5 2</v>
%       </property>
%     </object>
%   </property>
% \end{lstlisting}
An object may have properties which are again objects themselves (and which, in turn, may have further subobjects, etc.).
Again the object types are identified via the name of that property of the parent object.
Here \texttt{TRIANGULATION} is a \texttt{SimplicialComplex}.
This mechanism allows for rather elaborate constructions.

The maximal cells of the triangulation (called \texttt{FACETS}) are specified as subsets of the vertices of the polytope.
Each number refers to the corresponding row of the property \texttt{VERTICES} of the parent object.

Notice that a convex polytope can be triangulated in more than one way.
Therefore, \texttt{TRIANGULATION} is a property of a \texttt{Polytope} object which may contain several objects (of type \texttt{SimplicialComplex}).
The various triangulations are distinguished by their unique names.
These can be set by the user or are generated automatically (like here).

%------------------------------------------------------------

\section{Format specification in \texttt{RELAX NG}}
\label{sec:relax-ng}

The features presented above only provide a partial view of what can be expressed in \polymake's XML.
The full formal specification is expressed via \texttt{RELAX NG} (or \RNG for short) \cite{spec:RELAXNG}; see the Listings \ref{lstformat1} and \ref{lstformat2} below.
\RNG is a rather simple XML schema language based on the theory of hedge automata \cite{spec:HedgeAutomata}.
Table \ref{tablesyntax} contains a short overview of the compact \RNG syntax \cite{spec:RELAXNGcompact}.
The full specification file, which complies with the official \RNG standard and contains some additional explanatory annotations can be found in any current \polymake distribution under \url{[polymake_folder]/xml/datafile.rng}.

{\footnotesize
\setlength{\tabcolsep}{10pt}
\centering
\begin{table}
\begin{tabular}{|l l|}\hline
 \keyone{start }\ttfamily  = ... & Defines the pattern for the root element.\\
 \keythree{PatternName }\ttfamily = ... & Defines a pattern with a chosen name. \\
 \keyone{element},\keyone{ attribute} & Define XML tags / attributes.\\
 \ttfamily\{ ... \} & Describes the content of an element or attribute.\\
 \ttfamily | & Pattern alternatives. \\
 \ttfamily \& & Combine two patterns in arbitrary order.\\
 \ttfamily ?,+,* & Quantifiers: At most one, one or more, zero or more.\\\hline
\end{tabular}
\caption{\texttt{RELAX NG} compact syntax}\label{tablesyntax}
\end{table}
}

Listing \ref{lstformat1} contains pattern definitions for the higher level elements in \polymake's XML.
Each file either contains exactly one \texttt{object} or one \texttt{data} element as its root.
The pattern \texttt{ObjectContent} specifies that any object may contain multiple \texttt{property} and \texttt{attachment} elements.
Here we focus on objects, while loose data and attachments are discussed briefly at the end of this section.

\begin{lstlisting}[language=pmschema, frame=single,caption={Format specification, Part 1: top-level elements},label=lstformat1]
start = TopObject | LooseData

TopObject = element object { TopAttribs, ObjectContent }

TopAttribs = attribute type { 
    xsd:string { pattern = "[a-zA-Z][a-zA-Z_0-9]*::.*" } },
  attribute version { xsd:string { pattern = "[\d.]+" } }?,
  attribute tm { xsd:hexBinary }?

ObjectContent = 
  attribute name { text }?,
  attribute ext { text }?,
  element description { text }?,
  element credit { attribute product { text }, text }*,
  ( Property* & Attachment* )

Property = element property {
    SimpleName,
    attribute ext { text }?,
    ( ( attribute undef { "true" }, empty )
      | ( attribute type { text }?, PropertyData )
      | Text | SubObject+ ) }

SubObject = element object 
  { attribute type { text }?, ObjectContent }

Attachment = element attachment { 
  SimpleName, attribute ext { text }?, AttachmentData }

LooseData = element data {
  TopAttribs, attribute ext { text }?,
  element description { text }?, PropertyData }

SimpleName = attribute name 
  { xsd:string { pattern = "[a-zA-Z][a-zA-Z_0-9]*" } }
\end{lstlisting}

% minimally simplified by replacing PropertyName -> SimpleName in Propertty definition
% do not rename SimpleName in order to keep consistency

% PropertyName = attribute name { xsd:string      
%   { pattern="([a-zA-Z][a-zA-Z_0-9]*::)?[a-zA-Z][a-zA-Z_0-9]*"}}

Listing \ref{lstformat2} contains the pattern definitions for elements that encode actual content.
A \texttt{property} or \texttt{data} element can contain either a simple value stored in an attribute, a reference to another property, a container or a list of subobjects.
The \polymake XML format knows three container patterns: \texttt{Vector}, \texttt{Matrix} and \texttt{Tuple}.
The precise syntactical differences are somewhat subtle.
A \texttt{Matrix} is an array of containers of the same type.
In Example \ref{lstexample} the \texttt{VERTICES} (Lines 7--14) and the \texttt{FACETS} (Lines 15--23) of the square as well as the \texttt{FACETS} of the triangulation subobject (Lines 31--36) are matrices.
A \texttt{Vector} encodes an array of homogeneous content.
In  Example \ref{lstexample} the rows of the matrices mentioned above occur as vectors; additionally we have the \texttt{F\_VECTOR} of the triangulation (which counts the cells of the triangulation by dimension).
Sparse vectors employ the \texttt{e} element to specify the non-zero entries.
The final container pattern \texttt{Tuple} establishes records of heterogeneous content.
For maximum flexibility the three container types can be nested recursively.

\begin{lstlisting}[language=pmschema,firstnumber=39, frame=single,caption={Format specification, Part 2: Content elements}, label=lstformat2]
PropertyData = ( attribute value { text }, empty )
  | IdReference | Complex | element m { SubObject+ }

AttachmentData = 
  ( attribute type { text }?, attribute value { text }, empty )
  | ( attribute type { text }, attribute construct { text }?, 
    Complex ) | Text

Text = attribute type { "text" }, text

Complex = Vector | Matrix | Tuple

VectorContents = text
  | ( attribute dim { xsd:nonNegativeInteger }?,
      ( element e { ElementIndex, text }*
      | element t { ElementIndex?, TupleContents }+ ) )

ElementIndex = attribute i { xsd:nonNegativeInteger }

IdReference = element r { 
  attribute id { xsd:nonNegativeInteger }?, empty }

Vector = element v { VectorContents }

MatrixContents = 
  ( attribute cols { xsd:nonNegativeInteger }?, Vector* )
  | ( attribute dim { xsd:nonNegativeInteger },
      element v { ElementIndex, VectorContents }*)
  | Matrix+ | Tuple+

Matrix = element m { MatrixContents }

TupleContents = attribute id { xsd:nonNegativeInteger }?,
  ( text | ( Vector | Matrix | Tuple 
	    | IdReference | element e { text } )+ )

Tuple = element t { TupleContents }
\end{lstlisting}
%\section{Attachments and loose data}

\paragraph{Attachments.}%\label{attachments}

Attachments provide a mechanism for storing essentially arbitrary data with an object --- regardless of its type and the current version of \polymake.
They can be primitive data types as well as more complex types such as matrices and sets.
Object types such as \texttt{Polytope} are not allowed.
This can, for example, be used to store unrecognized data from pre-XML \polymake files or to keep track of relevant context data in an involved computation without having to create multiple files.
Every attachment is identified by a unique name.

\paragraph{Loose data.}%\label{loosedata}

The \polymake XML file format can also be used to store data which are not a full object.
The \texttt{object} node is then replaced by a \texttt{data} node and no properties or attachments may appear.
Otherwise, the format is essentially the same.
Listing \ref{lstloose} encodes an array whose single entry is the polynomial $\sqrt{5}/5 x^2 - y^3$.
Since the coefficients lie in the quadratic field extension $\Q(\sqrt{5})$, each of them is encoded as a \texttt{Tuple} $(a,b,c)$ which is to be read as $a+b\cdot\sqrt{c}$.
The polynomial again is a \texttt{Tuple} where the first entry encodes the terms (which form a \texttt{Matrix} of \texttt{Tuple} elements) and the second one the names of the variables.
%\begin{figure}
%<data type="common::Array&lt;Polynomial&lt;QuadraticExtension&lt;Rational&gt;, Int&gt;&gt;" version="3.0.2" xmlns="http://www.math.tu-berlin.de/polymake/#3">
\begin{lstlisting}[language=pmxml,caption={A file representing an array which contains one polynomial.},label=lstloose]
<data type="Array&lt;Polynomial&lt;QuadraticExtension&gt;&gt;"
      version="3.0"
      xmlns="http://www.math.tu-berlin.de/polymake/#3">
  <v>
    <t>
      <m>
        <t>
          <v dim="2"> <e i="0">2</e> </v>
          <t>0 1/5 5</t>
        </t>
        <t>
          <v dim="2"> <e i="1">3</e> </v>
          <t>-1 0 0</t>
        </t>
      </m>
      <t id="1">
        <v>x y</v>
      </t>
    </t>
  </v>
</data>
\end{lstlisting}
%\end{figure}

\paragraph{Element references}

To avoid writing the same data multiple times, an element can be replaced by a reference tag \texttt{<r>}, which points to another element using an identification number.
This is useful, for example, when storing multiple polynomials which all share the same variable names.
An example of an element using the \texttt{id} attribute can be seen in Listing \ref{lstloose}, line 16.
By referencing to this id, e.g., one can express that another polynomial is contained in the same ring.

% \begin{lstlisting}[language=pmxml,caption={Referencing an element using its \texttt{id}.},label=lstref]
%   <t id="1">
%     <v>x</v>
%   </t>
%   ...
%   <r id="1" />
% \end{lstlisting}

\section{Concluding remarks}

A key design decision is that the \polymake \RNG schema does not restrict the types of objects and their properties in any way.
It provides a simple syntax to recursively structure mathematical data in terms of vectors, matrices and tuples as it occurs in computations.
The precise type information relies on the implementation of the \polymake version specified.
In this way \polymake can be extended easily by adding new objects, new properties and new types.
The long-term sustainability of the data relies on the extra flexibility which comes from XSLT transformation style sheets.

It should be emphasized that this file format is by no means a replacement of existing standards such as OpenMath or (Content) MathML.
While MathML focuses on the \emph{presentation} of mathematical content, OpenMath and Content MathML are comprehensive frameworks for defining the \emph{semantics} of arbitrary mathematical information.
The \polymake XML format aims at something more modest: It provides a simple mechanism for storing \emph{concrete} mathematical data in a well-structured manner which still allows for extensions and modifications without breaking the overall concept.

\polymake's release documentation at 
\begin{center}
\url{http://polymake.org/release_docs/3.0/}  
\end{center}
is automatically generated.
This contains the complete list of objects, properties and their types.
We intend to enhance the mechanism for the documentation generation to export this information again as \RNG schema files.
This will allow third party developers to access \polymake data without relying on our software.

%------------------------------------------------------------

%\bibliographystyle{amsplain}
%\bibliography{main}

%\providecommand{\bysame}{\leavevmode\hbox to3em{\hrulefill}\thinspace}
%\providecommand{\MR}{\relax\ifhmode\unskip\space\fi MR }
% \providecommand{\MRhref}[2]{%
%   \href{http://www.ams.org/mathscinet-getitem?mr=#1}{#2}
% }
%\providecommand{\href}[2]{#2}

\end{document}